\documentclass[
  twoside,
  twocolumn,
  10pt,
  floatfix,
  superscriptaddress,
  prb,
  citeautoscript,
]{revtex4-1}

\usepackage{amsmath}
\usepackage{amssymb}
\usepackage{times}
\usepackage{enumitem}
\usepackage{graphicx}
\usepackage{units}
\usepackage[acronym]{glossaries}
\usepackage{hyperref}

\graphicspath{{./figs/}}

\newcommand{\eq}[1]{Eq.~(\ref{#1})}
\newcommand{\fig}[1]{Fig.~\ref{#1}}
\newcommand{\sect}[1]{Sect.~\ref{#1}}
\renewcommand{\vec}[1]{\ensuremath\boldsymbol{#1}}
\renewcommand{\epsilon}[0]{\varepsilon}
\newcommand{\icet}{\textsc{icet}}
\newcommand{\mchammer}{\textsc{mchammer}}
\newcommand{\sklearn}{\textsc{scikit-learn}}
\newcommand{\ase}{\textsc{ase}}
\newcommand{\spglib}{\textsc{spglib}}
\newcommand{\scipy}{\textsc{scipy}}
\newcommand{\basx}{Ba$_8$Al$_x$Si$_{46-x}$}

\newacronym{ardr}{ARDR}{automatic relevance detection regression}
\newacronym{ase}{ASE}{Atomic Simulation Environment}
\newacronym{bcc}{BCC}{body-centered cubic}
\newacronym{ce}{CE}{cluster expansion}
\newacronym{cs}{CS}{compressive sensing}
\newacronym{cv}{CV}{cross-validation}
\newacronym{dft}{DFT}{density functional theory}
\newacronym{eci}{ECI}{effective cluster interaction}
\newacronym{fcc}{FCC}{face-centered cubic}
\newacronym{lasso}{LASSO}{least absolute shrinkage and selection operator}
\newacronym{mc}{MC}{Monte Carlo}
\newacronym{ols}{OLS}{ordinary least-squares}
\newacronym{rfe}{RFE}{recursive feature elimination}
\newacronym{rmse}{RMSE}{root-mean-square error}
\newacronym{sgc}{SGC}{semi-grand canonical}
\newacronym{sof}{SOF}{site occupancy factor}
\newacronym{vcsgc}{VCSGC}{variance-constrained semi-grand canonical}
\newacronym{svd}{SVD}{singular value decomposition}

\setlist[itemize]{leftmargin=*, itemsep=0mm}
\setlist[enumerate]{leftmargin=*, itemsep=0mm}

\newcommand{\phys}{
  Chalmers University of Technology,
  Department of Physics,
  Gothenburg, Sweden
}
\newcommand{\ess}{
  Data Management and Software Centre,
  European Spallation Source,
  Copenhagen, Denmark
}

\begin{document}

\title{\icet{} -- A Python library for constructing and sampling alloy cluster expansions}

\author{Mattias {\AA}ngqvist}
\author{William A. Mu\~noz}
\author{J. Magnus Rahm}
\author{Erik Fransson}
\affiliation{\phys}
\author{C\'{e}line Durniak}
\author{Piotr Rozyczko}
\author{Thomas Holm Rod}
\affiliation{\ess}
\author{Paul Erhart}
\email{erhart@chalmers.se}
\affiliation{\phys}

\begin{abstract}
Alloy cluster expansions (CEs) provide an accurate and computationally efficient mapping of the potential energy surface of multi-component systems that enables comprehensive sampling of the many-dimensional configuration space.
Here, we introduce \icet{}, a flexible, extensible, and computationally efficient software package for the construction and sampling of CEs.
\icet{} is largely written in Python for easy integration in comprehensive workflows, including first-principles calculations for the generation of reference data and machine learning libraries for training and validation.
The package enables training using a variety of linear regression algorithms with and without regularization, Bayesian regression, feature selection, and cross-validation.
It also provides complementary functionality for structure enumeration and mapping as well as data management and analysis.
Potential applications are illustrated by two examples, including the computation of the phase diagram of a prototypical metallic alloy and the analysis of chemical ordering in an inorganic semiconductor.
\end{abstract}

\maketitle

\section{Introduction}

Ordering phenomena are ubiquitous in materials science, physics, and chemistry.
They are particularly relevant in multi-component systems, where they are associated for example with phase transitions, segregation as well as chemical order \cite{VanThoPuc18}.
The underlying energetics can usually be assessed by first-principles calculations, based on e.g., \gls{dft}, with good accuracy.
The computational cost of such calculations, however, precludes a statistically adequate sampling of the relevant configuration space.
In this context, alloy \glspl{ce} in combination with \gls{mc} simulations provide a powerful means to balance computational efficiency and accuracy \cite{SanDucGra84, Fon94}.

The \gls{ce} approach has been widely and very successfully adopted to analyze for example phase diagrams in metallic \cite{AstMcCFon93, OzoWolZun98a} and semiconducting alloys \cite{VanAydCed98, CedVenMar00, ZhoMaxCed06}, including surfaces \cite{DraReiFah01, SluKaw03, WelWieKer10, SteHamHwa10, CheSchSch11, CaoMue15, HerBraSch15} as well as nanoparticles \cite{MueCed10, CheButWal10, Yug11, Mue12, TanWanJoh12, WanTanJoh14, LiRacPu18, CaoLiMue18}.
\glspl{ce} have also been applied to study the temperature and composition dependence of ordering in various materials \cite{KimKavTho10, AngLinErh16, AngErh17, TroRigDra17, GunPucVan18}.
The \gls{ce} approach is not limited to the mapping of total and mixing energies but can also be applied to model for example activation barriers \cite{VanCedAst01}, vibrational properties \cite{MorWalCed00, WalCed02b}, chemical expansion \cite{AngErh17}, or transport properties \cite{AngLinErh16}.

Here, we introduce the integrated cluster expansion toolkit (\icet{}) to enable efficient construction and sampling of \glspl{ce}.
\textsc{icet} is designed to be modular, extensible, and flexible, while maintaining high computational efficiency.
This enables integration of \icet{} in extended workflows, reflecting the ongoing shift toward machine learning and large data initiatives in computational material science \cite{TayRosRoh14, PizCepSab16}.
\icet{} is primarily developed in Python whereas computationally more demanding parts are written in C++, providing performance while maintaining portability and ease-of-use.
This approach enables easy integration for example with countless first-principles codes and analysis tools accessible via the atomic simulation environment (\ase{}) \cite{LarMorBlo17} as well as state-of-the-art regression techniques via \sklearn{} \cite{PedVarGra11}.

\icet{} provides a feature set that is comparable to or extends beyond earlier monolithic codes, such as the ATAT \cite{WalCed02a, Wal09, WalAst02}, the UNCLE \cite{LerWieHar09} or CASM codes \cite{casm}.
Since \icet{} is written in Python, it is, however, straightforward to add new functionality.
This enables one for example to implement advanced algorithms for finding ground states \cite{HuaKitDac16, LarJacSch18}.
\icet{} is available under an open-source license and hosted on \textsc{gitlab} to encourage community participation.
Current functionality includes for example:
\begin{itemize}
\item
    support for multiple species and multiple coexisting sub-lattices, e.g., Ba$_{8-x}$Sr$_x$Ga$_y$Ge$_{46-y}$ or Au$_{1-x}$Pd$_x$:H$_y$
\item
    advanced linear regression techniques with regularization (including compressive sensing \cite{NelHarZho13}), cross-validation, and ensemble optimization via \sklearn{} \cite{PedVarGra11}
\item
    \gls{mc} simulations in various ensembles using observers and multiple \glspl{ce} in parallel via the \mchammer{} module
\item
    supplemental functionality including e.g., structure en\-um\-er\-a\-tion \cite{HarFor08, HarFor09}, structure mapping, convex hull ex\-tra\-ct\-ion \cite{BarBraDob96}, and ground state finding \cite{LarJacSch18}
\end{itemize}

The remainder of this paper is organized as follows.
The next section describes the methodologies implemented in \icet{}, including an overview of the \gls{ce} formalism, algorithms available for \gls{ce} construction, and the \mchammer{} module for sampling \glspl{ce} via \gls{mc} simulations.
The components and workflow of \icet{} are summarized in \sect{sect:workflow}.
Section~\ref{sect:applications} demonstrates the potential of \icet{} via examples.
The first example addresses the construction and sampling of \glspl{ce} for the Ag--Pd system as well as the subsequent generation of a phase diagram from these data.
The second example summarizes the application of \icet{} for the simulation of chemical ordering in a semiconducting system, specifically an inorganic clathrate.

\section{Cluster expansion formalism}
\label{sect:ce-formalism}

\subsection{Clusters and orbits}

\begin{figure*}
    \centering
    \includegraphics[width=0.95\linewidth]{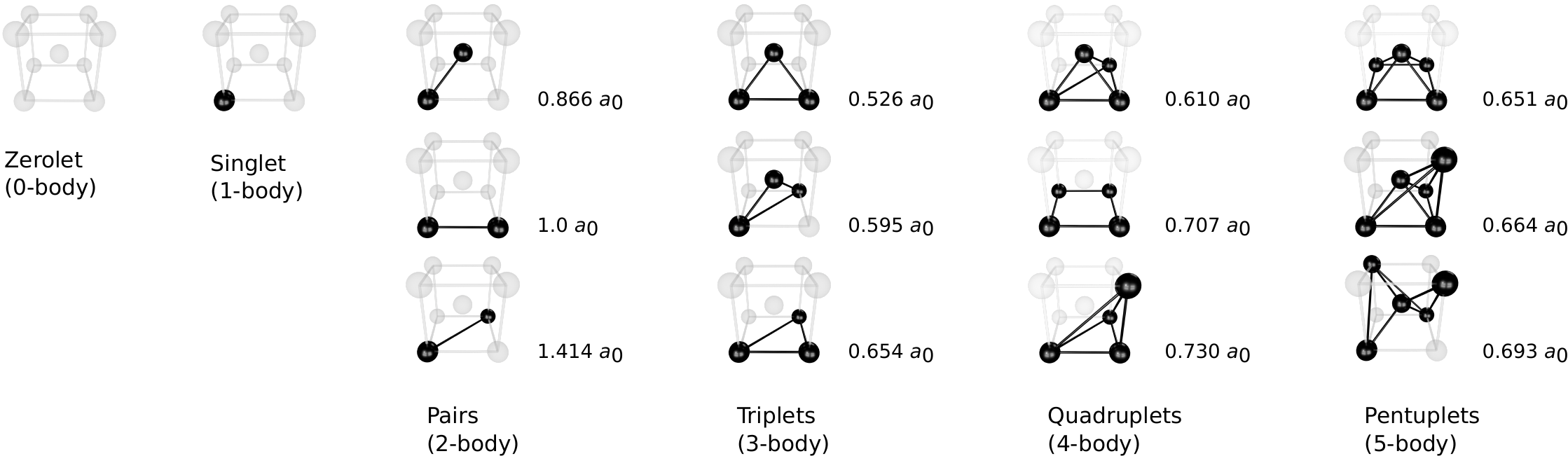}
    \caption{
        Illustration of clusters on a \gls{bcc} lattice.
        The clusters are ordered by radius from top to bottom, with the size given in units of the lattice constant $a_0$.
        The cluster radius is defined as the average distance of the sites to the center of the cluster.
    }
    \label{fig:first_clusters_in_bcc}
\end{figure*}

The objective of a \gls{ce} is to describe the variation of a property of interest, most commonly the energy, with the chemical configuration, i.e. the distribution of different species over a lattice.
To this end, the structure is decomposed into a set of clusters, where a cluster of order $k$ is defined as a list of $k$ sites.
Clusters are commonly referred to by order as singlets, pairs, triplets and so on.

\begin{figure}[b]
    \centering
    \includegraphics[width=0.8\linewidth]{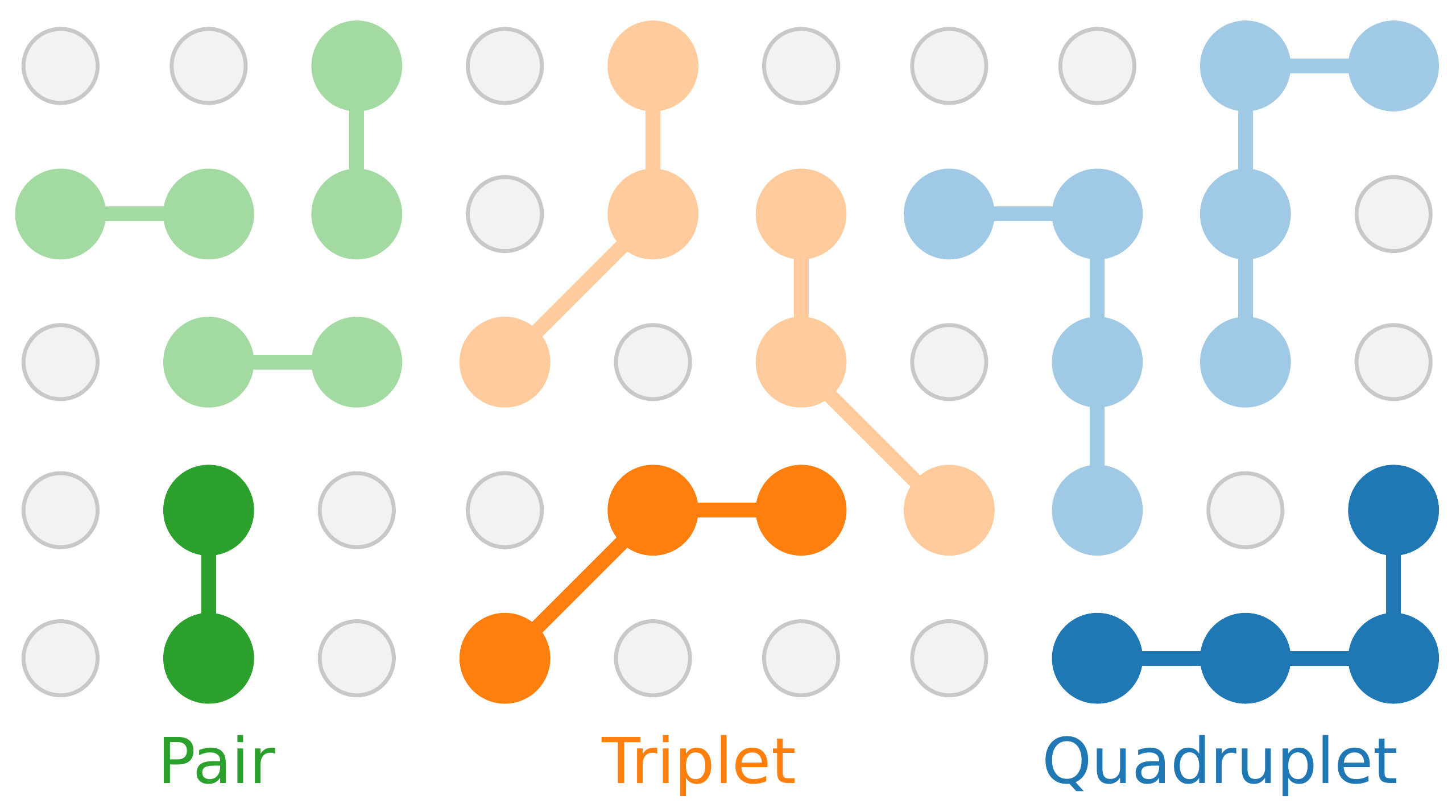}
    \caption{
        Examples for pair (green), triplet (orange) and quadruplet (blue) clusters on a square lattice.
        The representative (symmetry inequivalent) clusters are shown in dark colors, whereas examples for other clusters in the orbits are shaded.
    }
    \label{fig:symmetry-operations}
\end{figure}

The clusters are subject to the symmetry of the underlying lattice, as described by the associated space group (for periodic systems) or point group (for non-extended systems).
Clusters that can be mapped onto each other as a result of the application of an intrinsic symmetry operation of the lattice, are said to belong to the same orbit (\fig{fig:symmetry-operations}).
Each orbit in turn is represented by a symmetry inequivalent cluster, from which every other cluster of the orbit can be generated by application of the symmetry operations and a permutation of the sites.

While the number of clusters is in principle infinite, the short-ranged nature of physical interactions implies that clusters with notable contributions to the \gls{ce} are commonly shorter-ranged and few-bodied.
It is therefore customary to only include clusters up a certain size and order, as shown for the case of a \gls{bcc} lattice in \fig{fig:first_clusters_in_bcc}.

\subsection{Point functions}

It can be formally shown that a \gls{ce} is able to represent any function of the configuration $Q(\vec{\sigma})$ if one can construct a complete orthogonal basis \cite{SanDucGra84}.
Here, $\boldsymbol{\sigma}$ denotes the occupation vector, the $N$ elements of which indicate the species that resides on the corresponding site.
To obtain a practical procedure, for each lattice point $p$ one can define $M$ orthogonal point functions $\Theta_n(\sigma_p)$
\begin{align*}
    \Theta_{n}(\sigma_p)
    &= \begin{cases}
        1
        & \qquad \text{if $n=0$} \\
        -\cos\left(\pi(n+1)\sigma_p/M\right)
        & \qquad \text{if $n$ is odd} \\
        -\sin\left(\pi n \sigma_p/M\right)
        & \qquad \text{if $n$ is even},
    \end{cases}
\end{align*}
where $M$ is the allowed number of species and $n$ goes from 0 to $M-1$.
These point functions form an orthogonal set over all possible occupation numbers \cite{Wal09},
\begin{align*}
    \begin{aligned}
        \left \langle \Theta_n,\Theta_{n'}\right \rangle = \sum_{\sigma_p=0}^{M-1} \Theta_n(\sigma_p)\Theta_{n'}(\sigma_p) = \bigg \{ \begin{matrix}
            0 & \text{if $n \neq n'$}\\
            \neq 0 & \text{if $n = n'$}\\
        \end{matrix}
    \end{aligned}.
\end{align*}
In the case of multiple sublattices $M$ can assume different values for different sublattices.
For example, in the case of the zincblende alloy Al$_{1-x-y}$Ga$_x$In$_y$As$_{1-z}$Sb$_z$, $M=3$ for the cation and $M=2$ for the anion lattice.


\subsection{\texorpdfstring{\gls{ce}}{CE} expression}

A set of functions $\Pi_{\alpha}(\vec{\sigma})$ in the $M^N$-dimensional configuration space can now be constructed as products of point functions,
\begin{align*}
    \Pi_{\vec{\alpha}}(\vec{\sigma})
    &=
    \Theta_{n_1}(\sigma_1) \Theta_{n_2}(\sigma_2) \ldots
    \Theta_{n_l}(\sigma_l).
\end{align*}
Here, $\vec{\alpha} = [n_1, n_2,\ldots n_l]$, where $n_i$ are point function indices and $l$ is the number of sites in the structure.
Each $\vec{\alpha}$ corresponds to a cluster in the sense that $n_i = 0$ if site $i$ is not part of the cluster (such that the corresponding point function is $1$) and nonzero otherwise.
In the binary case, there is one $\vec{\alpha}$ for each cluster, each having elements that are either 0 or 1.
In systems with more than two components, there are multiple $\vec{\alpha}$ corresponding to the same cluster, due to the the possible combinations of nonzero point function indices $n_i$.

It can be shown that the functions $\Pi_{\vec{\alpha}}(\vec{\sigma})$ form a complete orthogonal set,
so that any function of the configuration can be expressed as
\begin{align*}
    Q(\vec{\sigma})
    &=
    \sum_{\vec{\alpha}} J_{\vec{\alpha}} \Pi_{\vec{\alpha}}(\vec{\sigma}).
\end{align*}
All basis functions $\Pi_{\vec{\alpha}}$ have one configuration invariant component that is equal to 1 when $\vec{\alpha}=\mathbf{0}$.
One can therefore omit the latter from the basis functions and instead use one configuration invariant term $Q_0$.
Furthermore, taking into account the symmetry of the clusters the summation can be carried out over orbits (or representative clusters) rather than all clusters, which yields the full \gls{ce} expression
\begin{align}
    Q(\vec{\sigma})
    &=
    Q_0 + \sum_{\vec{\alpha}}
    \left<\Pi_{\vec{\alpha}'}(\vec{\sigma}) \right>_{\vec{\alpha}}
    m_{\vec{\alpha}} J_{\vec{\alpha}}.
    \label{eq:cluster-expansion}
\end{align}
The $\left<\ldots\right>_{\vec{\alpha}}$ bracket indicates the average over the basis functions for all $\vec{\alpha}'$ that are symmetry equivalent to $\vec{\alpha}$.
The \glspl{eci} $J_{\vec{\alpha}}$ are the free parameters of the \gls{ce} and the target of the training procedure described in the next section.
Finally, $m_{\vec{\alpha}}$ denotes the multiplicity of the representative cluster $\vec{\alpha}$.

\section{Cluster expansion construction}
\label{sect:ce-construction}

\subsection{Matrix form}

To obtain a \gls{ce} for a specific material one must determine the \glspl{eci}.
To this end, one requires reference data in the form of a set of configurations $\{\vec{\sigma}_1, \vec{\sigma}_2, \ldots, \vec{\sigma}_n\}$ as well as an associated vector of target data $\vec{Q}^T = [Q_1, Q_2, \ldots Q_n]$, which is usually obtained from first-principles calculations.
Equation~\eqref{eq:cluster-expansion} can be cast in the form
\begin{align}
   \vec{Q} &= \vec{\Pi} \vec{J},
   \label{eq:cluster-expansion-matrix-form}
\end{align}
where the rows of $\vec{\Pi}$ are given by
\begin{align}
    \vec{\Pi}_i
    &= \left[ 1,
    \left<\Pi_{\vec{\alpha}'_1}(\vec{\sigma}_i) \right >_{\vec{\alpha}_1}m_{\vec{\alpha}_1},
    \right.
    \left.
    \ldots
    ,
    \left< \Pi_{\vec{\alpha}'_n}(\vec{\sigma}_i) \right>_{\vec{\alpha}_n}m_{\vec{\alpha}_n}
    \right ],
    \label{eq:cluster-vector}
\end{align}
and $\vec{J}$ denotes the \glspl{eci} with $J_0 = Q_0$.
Note that it can sometimes be useful to exclude $m_{\vec{\alpha}}$ from $\vec{\Pi}$ and let the target values $\vec{Q}$ refer to the primitive unit cell.
This will ensure all elements in $\vec{\Pi}$ are in the interval $[-1, 1]$ and avoids a bias due to the number of elements in $\vec{\sigma}$.

\subsection{Linear regression techniques}

Following the decomposition of a lattice into clusters and the construction of the sensing matrix, it remains to determine the \glspl{eci} by solving the linear system given by \eq{eq:cluster-expansion-matrix-form}.
This is equivalent to finding the parameter vector $\vec{J}$ that minimizes $\left\Vert\vec{\Pi}\vec{J} - \vec{Q}\right\Vert_2$.
In principle a solution can be determined by conventional least-squares, which works well in the overdetermined limit.
Due to the computational cost associated with \gls{dft} calculations the linear system is, however, often underdetermined and/or rows of the sensing matrix $\vec{\Pi}$ are correlated.
At the same time, the nearsightedness of physical interactions suggests that the solution vector $\vec{J}$ be sparse \cite{NelHarZho13}.
We are thus faced with the task of feature selection, a common process in machine learning, which yields models that are less prone to overfitting and more transferable.
It can also reduce the computational expense during sampling (see \sect{sect:ce-sampling}).

To achieve sparse solutions and combat overfitting one can include regularization terms in the objective function in the form of the $\ell_1$ or $\ell_2$-norm of the solution vector.
Consider, for example, elastic net regularization, for which
\begin{align*}
    \vec{J}_\text{opt}
    &=
    \underset{\vec{J}}{\text{argmin}}
    \left\{
        \Vert \vec{\Pi}\vec{J} - \vec{Q} \Vert^2_2
        + \alpha \Vert \vec{J} \Vert_1
        + \beta \Vert \vec{J} \Vert^2_2
    \right\}.
\end{align*}
For $\alpha=0$ this expression reduces to Ridge regression while for $\beta=0$ one obtains the expression commonly used for solving the \gls{lasso} problem.
Other techniques include the split-Bregman algorithm, which has been previously used for constructing \glspl{ce} \cite{NelHarZho13}, as well as \gls{rfe}, which iteratively removes the weakest parameters from a model and can be applied to different objective functions.
Furthermore, there are Bayesian linear regression techniques such as \gls{ardr}, which provide probabilistic models of the regression problem at hand, and Bayesian compressive sampling \cite{NelOzoRee13}.
The parameters that are intrinsic to a regression algorithm, such as $\alpha$ in the case of \gls{lasso} or the allowed number of features in the case of \gls{rfe}, are known as hyper-parameters.

The performance of different models can be assessed by \gls{cv}.
To this end, the available reference data is split into training and validation sets.
The former is used to fit a model, whereas the latter is used to measure the predictive power of the model, usually via the \gls{rmse}
\begin{align*}
    \text{RMSE} &=
    \sqrt{\frac{1}{N_s} \sum_i \left(Q_i^\text{model} - Q_i^\text{target}\right)^2},
\end{align*}
where the summation extends over the $N_s$ structures comprising the validation set.
To reduce the statistical error, the \gls{rmse} is furthermore averaged over several different splits of the reference data.

\icet{} supports various regression techniques, including the ones named above, via the \sklearn{} machine learning library  \cite{PedVarGra11} and allows one to compute \gls{cv} scores in a number of different ways.
Since sensing matrix and target data are readily available via the Python interface, one can also interface directly with \sklearn{} or other machine learning libraries.
\icet{} also provides functionality for generating ensembles of models from a single sensing matrix.
Thereby it is possible to check the sensitivity and stability of more advanced prediction, as illustrated in the examples below.

\section{Cluster expansion sampling}
\label{sect:ce-sampling}

A \gls{ce} can be employed in a number of different ways, including finding ground states \cite{LarJacSch18}, but it is most commonly sampled using \gls{mc} simulations.
For this purpose, \icet{} includes the \mchammer{} module, which supports various thermodynamic ensembles and provides supplemental functionality for data management and analysis.

\Gls{mc} sampling is carried out using the Metropolis algorithm, in which a trial is accepted with probability
\begin{align*}
    \mathcal{P}
    &= \min \left\{1, \exp\left(-\beta \Delta \psi\right) \right\}.
\end{align*}
Here, $\beta = 1 / k_B T$ and $\Delta \psi$ is the change in the thermodynamic potential associated with the ensemble being sampled (excluding the entropy term).

In the case of the canonical ($\{N_i\}VT$) ensemble the thermodynamic potential equals the internal energy, $\psi=E$, and a trial step involves swapping the species of two sites (conserving the concentrations).

Especially when exploring phase diagrams it is often useful to sample along the concentration axes.
This can be achieved for example by using the \gls{sgc} ($N\{\Delta\mu_i\}VT$) ensemble.
In this case, the trial move involves only one site, for which the occupation is swapped and the underlying thermodynamic potential is $\psi = E - N \sum_{i=2} c_i \Delta \mu_i $, where $N$ is the total number of sites, $c_i$ is the concentration of species $i$, and $\Delta \mu_i = \mu_i - \mu_1$ the chemical potential difference of species $i$ relative to the first species.

In the \gls{sgc} ensemble the mapping from chemical potential difference to concentration is multi-valued in two-phase regions.
It therefore cannot be used to sample across miscibility gaps.
This shortcoming can be overcome by employing the \gls{vcsgc} ensemble \cite{SadErh12}, which uses the same trial move as the \gls{sgc} ensemble but for which $\psi = E +  N k_\text{B} T \bar{\kappa} ( c + \bar{\phi} / 2)^2$.
The intensive parameters $\bar{\phi}$ and $\bar{\kappa}$ constrain respectively the average and the fluctuation of the concentration.

The choice of ensemble is motivated by the characteristics of the system and the objective of the simulations.
The canonical ensemble conserves concentrations, which makes it ideal for studying systems at specific compositions, extracting structural order parameters or obtaining ground states by simulated annealing.
The \gls{sgc} and \gls{vcsgc} ensembles, on the other hand, allow the composition to be continuously varied and by extension the integration of the free energy.
In the \gls{sgc} ensemble, the concentration derivative of the canonical free energy is given (for simplicity for a binary system) by
\begin{align}
    \partial \Delta F / \partial c &= - N \Delta \mu.
     \label{eq:sgc-dFdc}
\end{align}
This relation is, however, useful only in single-phase regions of the phase diagram, where $\Delta \mu$ maps to one and only one concentration.
In \gls{sgc} simulations, multi-phase regions manifest themselves by discontinuous jumps between phase boundaries.
Such discontinuities, if carefully studied, can be exploited for tracking phase boundaries but hinder integration of the free energy \cite{WalAst02}.

In the \gls{vcsgc} ensemble, the canonical free energy derivative is (for a sufficiently large system) given by \cite{SadErh12}
\begin{align}
    \partial \Delta F / \partial c
    &=
    - 2 N k_B T \bar{\kappa} \left(
    \left< c \right> + \bar{\phi} / 2 \right),
    \label{eq:vcsgc-dFdc}
\end{align}
where $\left< c \right>$ is the average observed concentration.
Unlike the \gls{sgc} ensemble, the mapping between $\bar{\phi}$ and concentration is always one-to-one for a sufficiently large value of $\bar{\kappa}$, whence the free energy can be recovered by integration across multi-phase regions.
The phase diagram can then be constructed with standard techniques for free energy minimization.

\Gls{mc} simulations in both the \gls{sgc} and \gls{vcsgc} ensembles are illustrated in \sect{sect:example-agpd} whereas the canonical ensemble is employed in \sect{sect:example-clathrate}.

\section{Workflow}
\label{sect:workflow}

\begin{figure}[b]
    \centering
    \includegraphics[width=0.98\columnwidth]{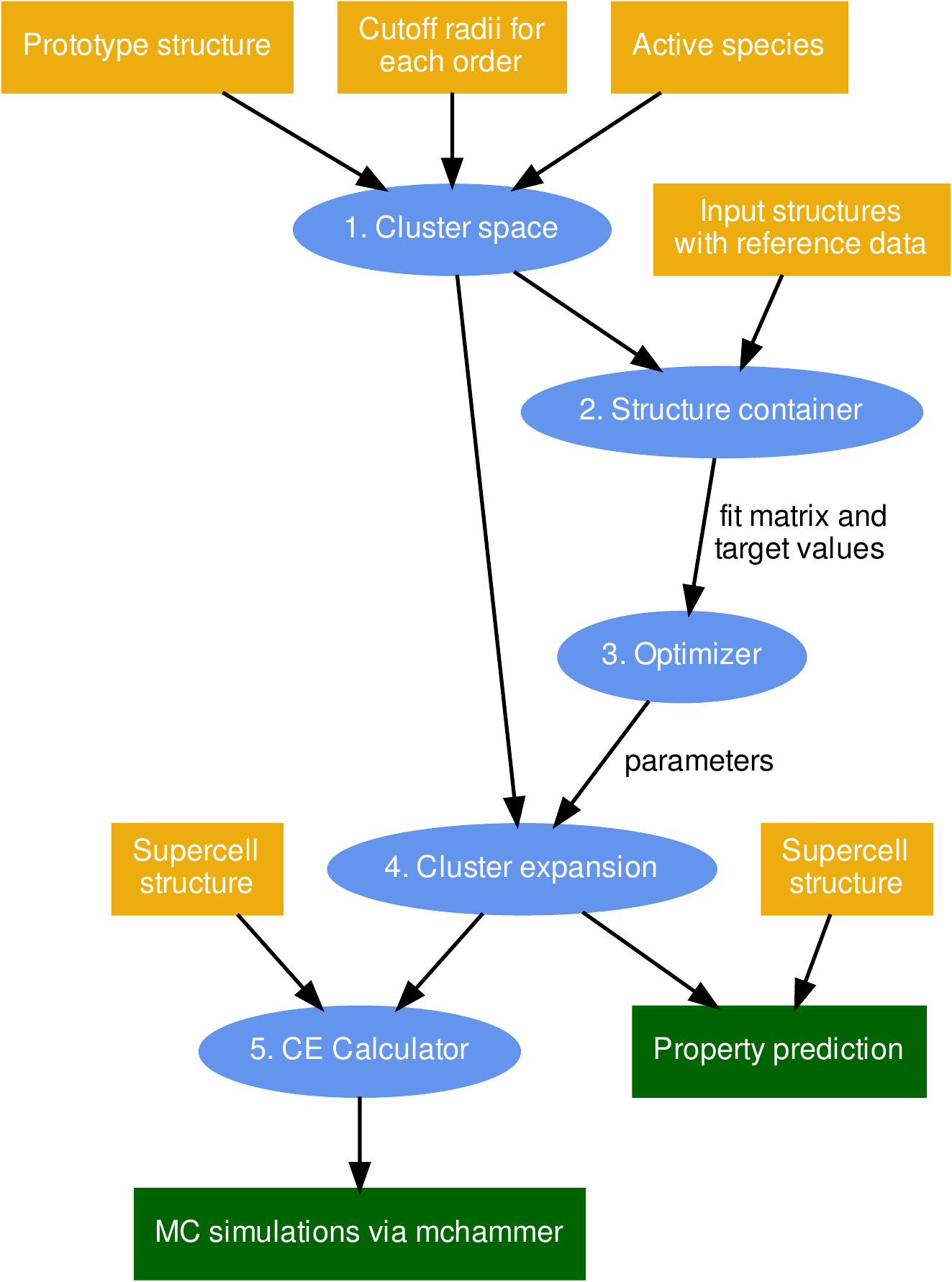}
    \caption{
        Illustration of the \icet{} workflow.
        Entities represented by Python objects are shown in blue, input parameters and data in orange, and additional functionalities in green.
    }
    \label{fig:workflow}
\end{figure}

\begin{figure}
    \centering
    \includegraphics{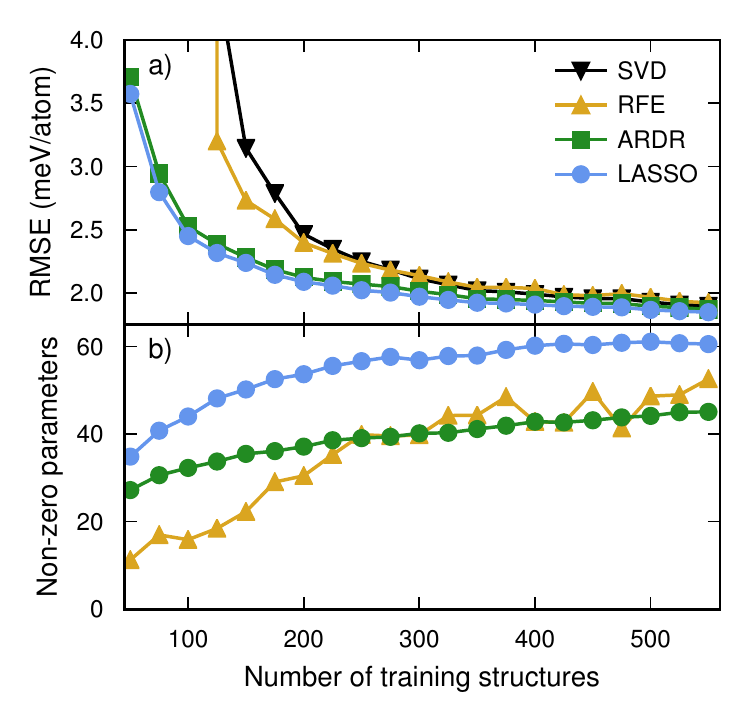}
    \includegraphics{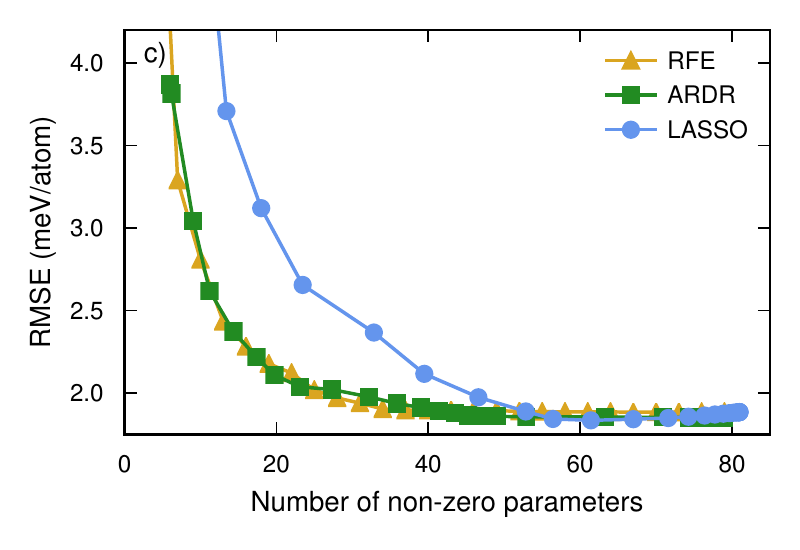}
    \caption{
        Construction of \glspl{ce} for Ag--Pd.
        (a) \Gls{rmse} obtained by \gls{cv} and (b) number of features (non-zero parameters) as a function of the number of training structures.
        (c) \Gls{cv}-\gls{rmse} as a function of the number of features using a training set size of 563 structures.
    }
    \label{fig:Ag_Pd_fitting}
\end{figure}

\icet{} integrates the \gls{ce} formalism (\sect{sect:ce-formalism}) with linear regression (\sect{sect:ce-construction}) and sampling techniques (\sect{sect:ce-sampling}) into one workflow (\fig{fig:workflow}).
Assuming reference data, say from \gls{dft} calculations, is available for a set of structures that have been generated for example by enumeration \cite{HarFor08} the construction and sampling of a \gls{ce} proceeds as follows.
\begin{enumerate}
\item
    The first step involves constructing a cluster space, which requires a prototype structure, a set of cutoff radii that define which clusters to include in the expansion, and a specification as to which species are allowed on each site.
    Internally, the \spglib{} library \cite{TogTan18} is employed to find the symmetries of the underlying lattice.
\item
    Next, cluster vectors are computed according to \eq{eq:cluster-vector} for all structures in the reference set and compiled into a structure container, which holds the sensing matrix $\vec{\Pi}$ as well as the vector of target values $\vec{Q}$.
\item
    Then linear regression techniques in combination with \gls{cv} are employed to solve \eq{eq:cluster-expansion-matrix-form} and find an optimal parameter vector $\vec{J}$.
    Here, externally provided optimization algorithms from, e.g., \sklearn{} and \scipy{} can be used.
\item
    Parameters and cluster space are subsequently combined to obtain the actual \gls{ce}, which can be used to predict the property in question for arbitrary supercells of the prototype structure.
\item
    For efficient sampling one can also set up a \gls{ce} calculator for a specific supercell to be used in \gls{mc} simulations via the \mchammer{} module.
\end{enumerate}

This workflow is supplemented by a number of tools, which enable one for example to enumerate structures \cite{HarFor08}, extract the convex hull, map relaxed structures onto ideal lattices, or find ground states \cite{LarJacSch18}.

A full description of the different entities involved in this process and the Python objects that describe them can be found in the \icet{} user guide \cite{icet_user_guide}.

\section{Applications}
\label{sect:applications}

\subsection{Phase diagram of the Ag--Pd system}
\label{sect:example-agpd}

\subsubsection{Reference calculations}

As a first example for the application of \icet{}, we describe the construction of a \gls{ce} for the \gls{fcc} Ag--Pd alloy.
To this end, a database of 631 reference structures corresponding to all distinct supercells with up to 8 atoms was set up using the structure enumeration feature of \icet{}.
More refined structure selection approaches are available \cite{NelHarZho13} but are not considered in this example.

Reference calculations were then carried out using \gls{dft} calculations using the projector augmented method \cite{Blo94, KreJou99} as implemented in \textsc{vasp} \cite{KreFur96a, KreFur96b}.
The van-der-Waals density functional method \cite{DioRydSch04, KliBowMic11} with consistent exchange \cite{BerHyl14}, which has been shown to be very well suited for transition metals \cite{GhaErhHyl17}, was employed to represent the exchange-correlation potential.
The Brillouin zone was sampled using $\vec{k}$-point grids equivalent to a $18\times18\times18$-mesh for the primitive \gls{fcc} unit cell and the plane wave cutoff energy was set to \unit[384]{eV}.
Both the atomic positions and the cell metric were relaxed until residual forces and stress were less than \unit[10]{meV/\AA{}} and \unit[0.8]{GPa}, respectively.
Relaxations were carried out using first-order Methfessel-Paxton smearing with a width of \unit[0.1]{eV}.
Final energy calculations were carried out using for the relaxed structures the tetrahedron method with Bl\"ochl corrections using a smearing width of \unit[0.05]{eV}.

\subsubsection{Construction of \texorpdfstring{\gls{ce}}{CE} models}

\begin{figure}
    \centering
    \includegraphics[scale=0.9]{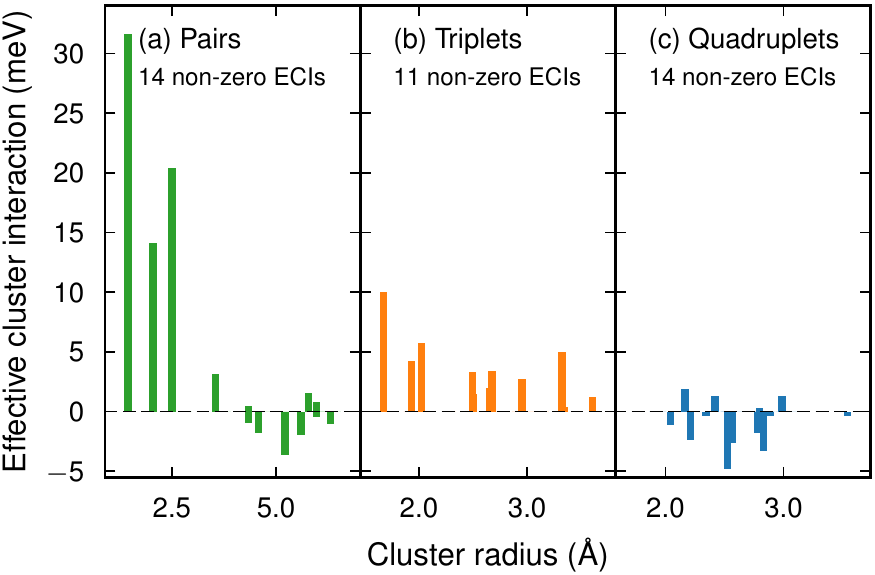}
    \caption{
        \Glspl{eci} for the Ag--Pd system obtained using the \gls{ardr} method with all 631 structures in the training set.
        The \glspl{eci} illustrate the sparseness of the solution and reflect the short-ranged nature of the interactions.
        The cluster radius is defined as the mean distance from the center of the cluster to any of its points.
    }
    \label{fig:Ag-Pd-ecis}
\end{figure}

Next a cluster space was set up, including clusters up to fourth order with cutoffs of \unit[3.3]{$a_0$}, \unit[1.6]{$a_0$}, and \unit[1.5]{$a_0$} in units of the lattice parameter $a_0$ for pairs, triplets, and quadruplets, respectively.
The resulting cluster space contained 81 parameters, including 1 zerolet, 1 singlet, 24 pairs, 20 triplets, and 35 quadruplets.
In principle, \icet{} does not impose limits on cluster order or size.

In order to study the convergence with respect to the number of training structures we computed the learning curve using the \gls{rmse} score averaged over the validation set using the \gls{cv} estimator functionality in \icet{}.
The latter generates an estimate for the \gls{cv}-\gls{rmse} score by using the shuffle-split method with 50 splits, while the final \gls{ce} is obtained by training against the complete data set.
Four different optimization methods were compared, including singular value decomposition, \gls{lasso}, \gls{rfe} based on \gls{ols}, and \gls{ardr} (\fig{fig:Ag_Pd_fitting}a).

\begin{figure}
    \centering
    \includegraphics[width=0.88\columnwidth]{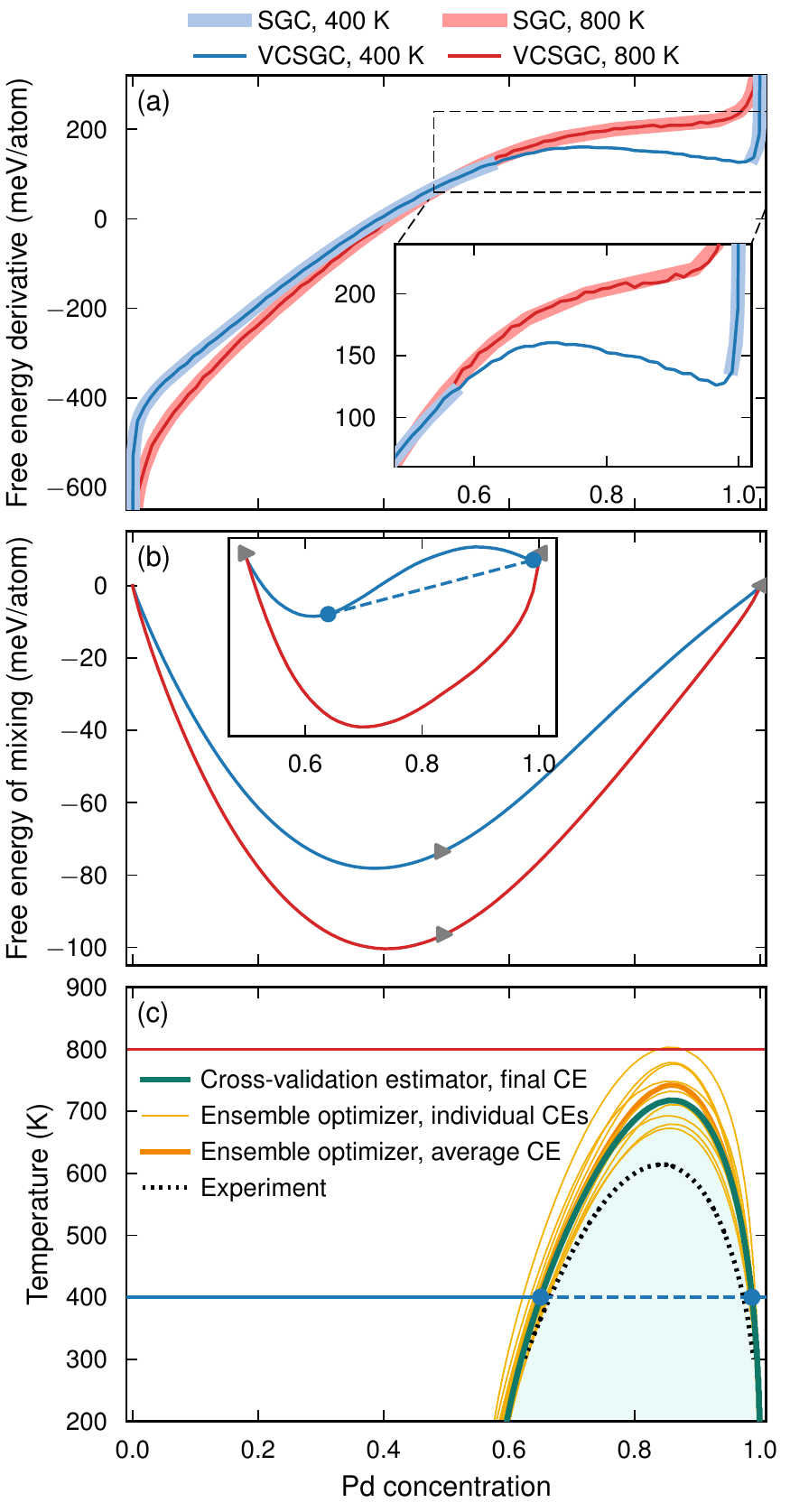}
    \caption{
        Ag--Pd system.
        (a) Free energy derivative $\partial F / \partial c$ obtained by \gls{mc} sampling in the \gls{sgc} and \gls{vcsgc} ensembles.
        The \gls{sgc} ensemble cannot yield stable solutions in the two-phase region (inset).
        (b) Free energy of mixing obtained by numerical integration in the \gls{vcsgc} ensemble.
        The inset indicates the behavior between $c=0.5$ and $1.0$ where the free energy is concave at low temperature.
        (c) Phase diagram constructed from the free energy landscape using \glspl{ce} obtained using the \gls{cv} estimator (green line) as well as the ensemble optimizer (orange lines).
        Experimental data (dotted black line) from Ref.~\onlinecite{DinWatKro08} obtained via the CALPHAD method.
    }
    \label{fig:phase-diagram}
\end{figure}

In the overdetermined region all methods yield similar \gls{cv}-\gls{rmse} scores; in the underdetermined region, though, \gls{lasso} and \gls{ardr} outperform the other two methods.
While \gls{lasso} and \gls{ardr} have almost identical \gls{cv} scores there is, however, a significant difference in the number of features, i.e. non-zero parameters (\fig{fig:Ag_Pd_fitting}b).
To study this behavior further we analyzed the \gls{cv}-\gls{rmse} score as a function of the number of features in the solution (\fig{fig:Ag_Pd_fitting}c).
To this end, we used a training set of 563 structures and scanned the hyper-parameters of the regression algorithms to control the sparsity of the solution.

This analysis reveals that \gls{ardr} and \gls{rfe} with \gls{ols} yield \glspl{ce} with a very low \gls{cv}-\gls{rmse} of \unit[2]{meV/atom} using only 30 features.
\gls{lasso} reaches the same level but yields about 50 features.
In this case, \gls{ardr} is thus the method of choice since it converges as quickly with the number of training structures as \gls{lasso} but yields a smaller number of orbits with non-zero \glspl{eci} (\fig{fig:Ag-Pd-ecis}), which leads to a more transferable model and reduces the computational cost for sampling.

\subsubsection{Phase diagram from \texorpdfstring{\gls{mc}}{MC} simulations}

To construct a phase diagram for the Ag--Pd system, we employed the \mchammer{} module for sampling a $5\times5\times5$ conventional \gls{fcc} supercell (500 sites) in both the \gls{sgc} and \gls{vcsgc} ensembles.
We carried out 200,000 \gls{mc} trial steps with the \gls{sgc} ensemble at 105 values of $\Delta \mu$ in the range $[-1.04, 1.04]$ and the same number of trial steps with the \gls{vcsgc} ensemble at $\bar{\kappa}=200$ and 105 values of $\bar{\phi}$ in the range $[-2.3, 0.3]$.
Simulations were run at \unit[25]{K} intervals between 100 and \unit[900]{K}, corresponding in total to approximately $6.9 \times 10^8$ trial steps or $1.4 \times 10^6$ \gls{mc} cycles.

Simulations were first carried out for the \gls{ce} described above that was constructed using the \gls{cv} estimator with \gls{ardr} and the full dataset of 631 structures (\gls{cv}-\gls{rmse} \unit[2]{meV/atom}).

The free energy derivatives extracted from the \gls{sgc} and \gls{vcsgc} simulations (using Eqs.~\eqref{eq:sgc-dFdc} and \eqref{eq:vcsgc-dFdc}) coincide everywhere except for a region on the Pd-rich side at lower temperatures where the \gls{sgc} simulations exhibit a discontinuity, which is the hallmark of a two-phase region (\fig{fig:phase-diagram}a).
In this situation, the full free energy can thus only be recovered from the \gls{vcsgc} simulations (\fig{fig:phase-diagram}b).
The two-phase region is manifested by a concave region in the free energy of mixing (inset in \fig{fig:phase-diagram}b).

To construct the phase diagram, we fitted the free energy of mixing \fig{fig:phase-diagram}b) to third-order Redlich--Kister polynomials at each temperature separately and then fitted the temperature dependence of the polynomial expansion coefficients to conventional third-order polynomials.
This representation provides a continuous and smooth map of the free energy in both temperature and composition, akin to the CALPHAD approach \cite{DinWatKro08}.
The predicted phase diagram exhibits a pronounced miscibility gap on the Pd-rich side with a critical temperature $T_c$ of \unit[718]{K}.
This is overall in good agreement with the experimental result \cite{DinWatKro08}, except for an overestimation of $T_c$, which has been experimentally determined as approximately \unit[610]{K}.

To illustrate the sensitivity of the phase diagram to variations in the training set, we also considered a set of ten \glspl{ce} that were obtained using the \gls{ardr} method via the ensemble optimizer functionality of \icet{}.
The latter approach enables one to generate a series of \glspl{ce} that are trained in identical fashion but are based on different training sets.
The latter are generated by selection with replacement (bagging) from the reference data set such that the number of structures in the training sets equals the number of reference structures.
The thus obtained \glspl{ce} are numerically very similar to the \gls{ce} obtained using the \gls{cv} estimator approach before (\fig{fig:Ag_Pd_fitting}).
Using this set of \glspl{ce} enabled us to estimate not only the error in the mixing energies but its impact on the final phase diagram.

The \gls{ce} obtained by averaging the \glspl{eci} over all \glspl{ce} in the ensemble yields a $T_c$ of \unit[742]{K} that is only slightly higher than the \gls{ce} generated using the \gls{cv} estimator.
The individual \glspl{ce} yield, however, a larger variation, spanning the range from 673 to \unit[803]{K}.
This demonstrates that \gls{cv}-\gls{rmse} alone is an insufficient measure of the quality of a \gls{ce}.
Rather a more careful assessment of the quantity of interest must be carried out if an accurate estimate is required.

\subsection{Chemical ordering in an inorganic clathrate}
\label{sect:example-clathrate}

\subsubsection{Background and reference structures}

The \gls{ce} approach is not limited to metallic system and the prediction of phase diagrams.
It is also very useful to model for example the degree of ordering as a function of temperature and composition.
The latter can in turn can be experimentally assessed using diffraction techniques based on X-ray or neutron scattering \cite{ChrJohIve10}.
Here, this possibility is illustrated by using \icet{} to predict the \glspl{sof} in the inorganic clathrate \basx{}.

Clathrates are inclusion compounds with a lattice structure that can trap atomic or small molecular species.
\basx{} falls into the class of type-I clathrates, which belong to spacegroup Pm$\bar{3}$n.\cite{SheKov11}
In this case, the host lattice is made up of Al and Si atoms, which occupy Wyckoff sites $6c$, $16i$, and $24k$, whereas Ba atoms reside inside the cages for charge balance.
Al and Si do not occupy the framework sites randomly but exhibit some degree of ordering, which results from a delicate balance between energy and entropy and can be experimentally accessed via the \glspl{sof}.
While for a completely random distribution at, e.g.,  $x=12$ one would expect \glspl{sof} of 12/46=26\%\ for all sites, in \basx{} one observes values in the range from close to zero to 80\%\ \cite{RouCruCha12}.
Detailed studies of this behavior, including careful comparison with experiment, have been published elsewhere \cite{AngLinErh16, AngErh17, TroRigDra17}.
Here, we focus on the construction and sampling of \glspl{ce} for this system, in particular highlighting the analysis capabilities provided by \icet{}.

\subsubsection{\texorpdfstring{\gls{ce}}{CE} construction}

The unit cell contains 46 framework sites, which prevents an enumeration approach for structure generation.
Instead, 240 occupations of the primitive unit cell for $x=(13\ldots16)$ were produced by randomly distributing Al and Si atoms over the host lattice.
The structures were relaxed using \gls{dft} calculations using a similar procedure as for the Ag--Pd structures described above.
The PBE exchange-correlation functional was used \cite{PerBurErn96} with a plane wave energy cutoff of 319\,eV and a $\Gamma$-centered $3\times3\times3$ $\vec{k}$-point mesh.
The other parameters were identical to those given in \sect{sect:example-agpd}.

\begin{figure}
    \centering
    \includegraphics{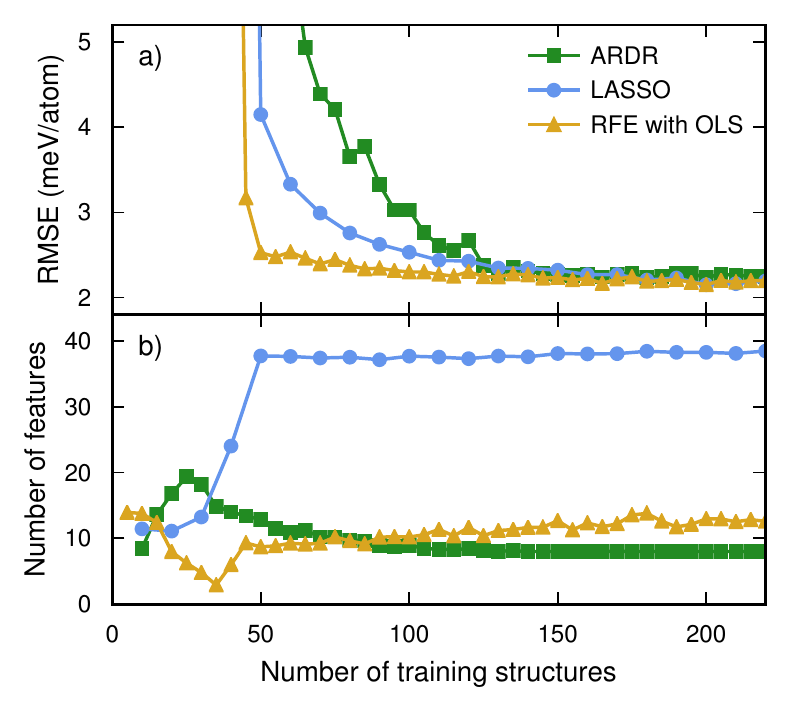}
    \includegraphics{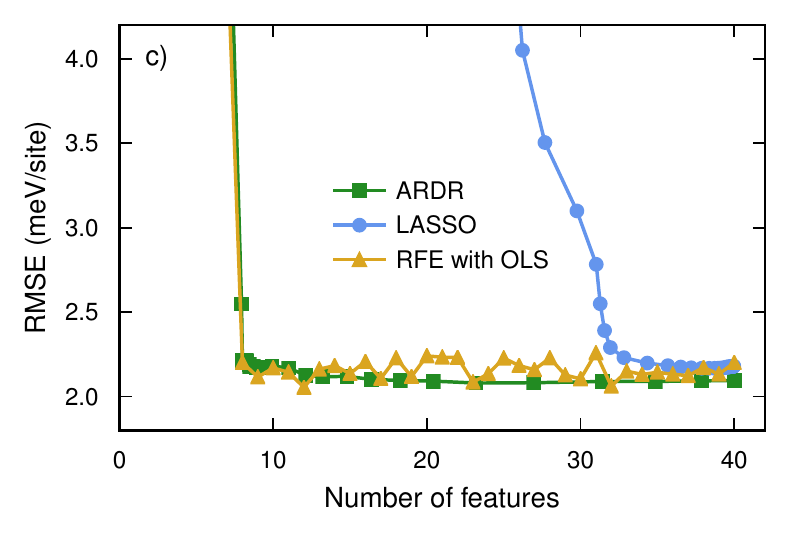}
    \caption{
        Construction of \glspl{ce} for \basx{}.
        (a) \gls{cv}-\gls{rmse} and (b) number of features as a function of the number of training structures.
        (c) \gls{cv}-\gls{rmse} as a function of the number of features obtained by varying the hyper-parameters of the respective methods using a fixed training set comprising 240 structures.
    }
    \label{fig:clathrate_learning_curves}
\end{figure}

A cluster basis was constructed using a cutoff of \unit[0.49]{$a_0$} for both pairs and triplets, resulting in 13 and 23 symmetry inequivalent clusters, respectively.
\Glspl{ce} were generated using \gls{ardr}, \gls{lasso}, and \gls{rfe} with \gls{ols} in conjunction with the \gls{cv} estimator functionality based on the shuffle-split method with 50 splits.

\Gls{rfe} with \gls{ols} yields both fast convergence with training set size and sparse solutions (\fig{fig:clathrate_learning_curves}a,b).
\Gls{lasso} and \gls{ardr} require almost twice as many training structures to achieve similar \gls{cv}-\gls{rmse} values.
\gls{ardr} provides sparse solutions that are similar to those obtained by \gls{rfe} with \gls{ols}.
The \gls{ce} models obtained by \gls{lasso}, however, have a much larger number of features (\fig{fig:clathrate_learning_curves}c).
These trends are similar to the case of Ag--Pd (\sect{sect:example-agpd}) except for the roles of \gls{ardr} and \gls{rfe} with \gls{ols} being reversed.

\subsubsection{\texorpdfstring{\gls{ce}}{CE} sampling}

To predict the \glspl{sof} as a function of temperature, we employed the \mchammer{} module for sampling a $2\times2\times2$ supercell (268 sites) in the canonical ensemble at $x=12$, for which experimental data is available \cite{RouCruCha12}, using a simulated annealing approach.
The temperature was decreased from 1200 to \unit[0]{K} at a rate of \unit[100]{K}/\unit[10,000]{MC cycles}, corresponding to a total of almost 48 million trial steps.
The \glspl{sof} were monitored during the simulation using the observer functionality of \icet{}, which enables one to compute various quantities of interest at specified intervals along the trajectory.

The simulations were first carried out using the \gls{ce} obtained using \gls{rfe} with \gls{ols} and a set of 240 structures.
To estimate the statistical reliability of the thus predicted \glspl{sof}, we furthermore ran the simulations for an ensemble of models that were generated from the available data (bagging), in almost identical fashion as for the Ag--Pd models described above.

\begin{figure}
    \centering
    \includegraphics{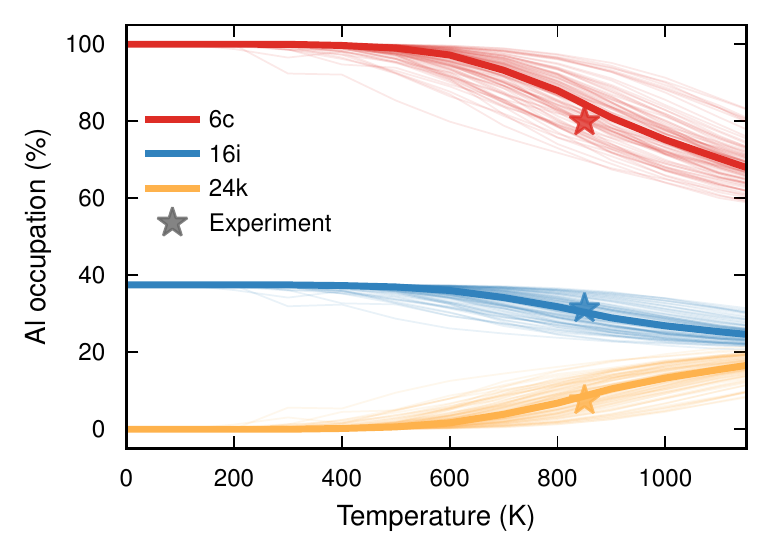}
    \caption{
        Al \glspl{sof} as a function of temperature in \basx{} with $x=12$ from simulated annealing.
        Simulations were carried out for an ensemble of \gls{ce} models obtained by different 90-10 splits of the reference data set (thin shaded lines) as well as the \gls{ce} obtained by averaging over the ensemble (bold line).
        Experimental data from Ref.~\onlinecite{RouCruCha12}.
    }
    \label{fig:clathrates_sofs}
\end{figure}

At the composition of $x=12$ the number of nearest neighbor Al--Al pairs is zero over the entire temperature range.
This behavior is due to Al--Al bonds being energetically unfavorable, which is  familiar from the Loewenstein rule for zeolites.
While the Al--Al pair distribution is experimentally practically impossible to access, diffraction experiments can provide information about the \glspl{sof}, which are ultimately the result of the interplay of energy and entropy \cite{ChrJohIve10}.
The \glspl{sof} obtained from \gls{mc} simulations exhibit a systematic variation with temperature and strongly deviate from the random limit, which would correspond to approximately 35\%\ (\fig{fig:clathrates_sofs}).
At temperatures below approximately \unit[600]{K} the \glspl{sof} converge and the structure adopts a well ordered configuration.
The ground state at this stoichiometry corresponds to Al \glspl{sof} of 100\%, 37.5\%, and 0\%\ for Wyckoff sites 6c, 16i, and 24k, respectively, in agreement with earlier simulations \cite{AngErh17,TroRigDra17}.

The calculated \glspl{sof} compare very well with the available experimental data in the temperature interval between 800 and \unit[900]{K} \cite{RouCruCha12}.
This is very reasonable since it is likely that kinetic factors prevent full ordering in the experiments.
A more extensive analysis of the \glspl{sof} as both a function of temperature and composition \cite{AngErh17} shows very good comparison with experiment over the entire composition range \cite{ChrJohIve10, RouCruCha12}, which further validates the present approach.
The ground state configurations obtained in the ze\-ro-temp\-er\-at\-ure limit are valuable in themselves as they provide ordered prototypes, which can be used for further analyses of e.g., electrical \cite{AngLinErh16} and thermal transport properties \cite{LinBroFra19}.

\section{Conclusions}
\label{sect:conclusions}

In the present paper, we have introduced the \icet{} Python package for the construction and sampling of alloy \glspl{ce}.
Thanks to its modular design it can be readily extended and combined with other Python packages to achieve complex functionalities.
It thereby also provides an excellent environment for method development, including the adaptation of further machine learning techniques (e.g., genetic algorithms \cite{BluHarWal05}) and the integration in extended workflows, e.g., in the context of high-throughput computing \cite{TayRosRoh14, PizCepSab16}.
\icet{} also provides a number of supplementary features pertaining, e.g., to structure enumeration and mapping as well as data analysis.

\icet{} readily supports a variety of different methods for constructing \glspl{ce}, as demonstrated specifically for the metallic alloy Ag--Pd and the inorganic clathrate \basx{}.
Several different regression methods were compared, illustrating a balance between the sparsity of the solution and the accuracy of the final \gls{ce}.
While for Ag--Pd the \gls{ardr} method provided the best performance, yielding both low \gls{cv} scores and a sparse solution, \gls{rfe} based on \gls{ols} achieved the best results in the case of \basx{}.
Regression using \gls{lasso} led to less optimal solutions in both situations, an observation that we have also made in the case of force constant models \cite{EriFraErh19, FraEriErh19}.

\icet{} also allows construction of ensembles of \gls{ce} models, which provides a powerful means to investigate the sensitivity of a prediction to variations in the underlying model.
Specifically, this enables one to extrapolate the impact of the statistical uncertainty in \gls{ce} models to complex observables such as a phase diagram or \glspl{sof}.
In the case of the Ag--Pd system, this approach was for example adopted to demonstrate that a set of models with numerically similar \gls{cv} scores can yield variations in the critical temperature on the order of \unit[100]{K}, providing an estimate of average and standard deviation.
The same approach was employed for the \basx{} system to determine the statistical uncertainty of the predicted temperature dependence of the \glspl{sof}.

The \mchammer{} module of \icet{} includes functionality for extracting additional information from \gls{mc} trajectories, as illustrated by the application to the inorganic clathrate \basx{}.
This allows one to observe for example structural order parameters throughout a simulation, including \glspl{sof}, neighbor counts, or short-range order parameters \cite{Cow49}.

Overall \icet{} package is thus well suited for efficient construction and sampling of \glspl{ce}, for example in high-throughput schemes.
For such applications one must, however, not only consider the computational effort but also the amount of human intervention required.
In the future, it is therefore desirable to set up protocols that further automatize the selection of e.g., regression method, hyper-parameters, and training set size.

\icet{} is provided under an open-source license.
Its development is hosted on \textsc{gitlab} to encourage community participation and the most recent released version can be conveniently installed via \textsc{PyPi}.
A comprehensive user guide with an extensive tutorial section is available online \cite{icet_user_guide}.

\section*{Acknowledgments}
This work was funded by the Knut and Alice Wallen\-berg Foundation (KAW), the Swedish Research Council (VR), the Swedish Foundation for Strategic Research (SSF), and the Interreg 
programme of the  European Union via the MAX4ESSFUN subproject.
Com\-puter time allocations by the SNIC at C3SE (Gothenburg), NSC (Link\"oping), and PDC (Stockholm) are gratefully acknowledged.

\end{document}